# بهبود قابلیت اطمینان و عمرباتری در شبکه‌های ارتباطی اینترنت اشیا: چالش‌ها و راهبردهای مبتنی بر یادگیری ماشینی

امین آذری[1]، محمود عباسی[2]

۱- دانشگاه KTH، سوئد
– aazari@kth.se
۲- دانشگاه آزاد اسلامی مشهد – مشهد – ایران
– mahmoud_abbasi_student@yahoo.com

**چکیده:** در راستای تحقق یک جامعه‌ی هوشمند، برقراری ارتباط برای تمام اشیاء هوشمند با هزینه و انرژی مصرفی کم یک نیاز اساسی می‌باشد. در حالی که شبکه‌های بی‌سیم کنونی برای برقراری بهینه ارتباطات، نیاز به مدیریت متمرکز شبکه و منابع دارند، عواملی مانند انرژی مصرفی در ارسال-دریافت سیگنال‌های کنترلی و تعداد زیاد دستگاه‌های اینترنت اشیاء، امکان استفاده از چنین رویکردهای متمرکزی را در آینده غیرممکن خواهد ساخت. به منظور حل این مشکل، در این مقاله به بررسی امکان استفاده از راه‌حل‌های مبتنی بر یادگیری ماشینی برای شبکه‌های اینترنت اشیاء پرداخته‌ایم. در گام اول برای دستیابی به این هدف، روش‌های یادگیری با پیچیدگی کم، که مناسب پیاده‌سازی در اشیاء می‌باشند، مورد بررسی قرار گرفته‌اند. در ادامه، یک روش یادگیری برای تطبیق پارامترهای مخابراتی در اشیاء با محیط پیرامون آنها ارائه شده است. در این روش پیشنهادی، تابعِ ارزش هر تصمیم بر اساس سابقه‌ی انرژی مصرفی و میزان موفقیت در اتخاذ آن تصمیم طراحی می‌شود. این طراحی دستگاه را قادر می‌سازد تا بهترین مصالحه را بین انرژی مصرفی و قابلیت اطمینان ارتباطات بدست آورد. در گام بعدی، مقایسه عملکرد روش پیشنهادی مقاله با رویکرد سیستم‌های متمرکز، با بهره‌گیری از ابزار هندسه تصادفی، ارائه شده است. سپس، ارتباط بین پارامترهای الگوریتم یادگیری ماشینی و عملکرد سیستم مخابراتی، مانند انرژی مصرفی و قابلیت اطمینان، مورد تجزیه و تحلیل قرار گرفته است. نتایج شبیه‌سازی نشان می‌دهد که در مقایسه با جدیدترین روش‌های پیشنهاد شده در ادبیات این موضوع، هردو معیار بهره‌وری انرژی و سطح اطمینان می‌تواند با استفاده از روش یادگیری مندرج در این مقاله به صورت قابل توجهی بهبود یابند.

**واژه های کلیدی:** اینترنت اشیاء، عمر باتری، یادگیری، نسل پنجم، یادگیری ماشینی

## ۱- مقدمه

در طول پیدایش شبکه‌های مخابراتی بخش قابل توجهی از تحقیقات در این حوزه تمرکز خود را بر روی مقاوم‌سازی سیستم‌های ارتباطی در مقابل نویز و تداخل قرار داده‌اند که این مسئله عمدتاً ناشی از محدودیت‌های کانال فیزیکی مانند نویز و تداخل می‌باشد. به لطف پهنای باند بالا و استفاده از سخت‌افزارها و نرم‌افزارهای پیشرفته در نسل چهارم شبکه تلفن همراه (هم در ایستگاه‌های پایه و هم در دستگاه‌های ارتباطی کاربران)، این شبکه قادر به ارائه ارتباطات با سرعت بالا، به صورت یکپارچه و قابل اطمینان به کاربران می‌باشد [۱]. در مقایسه با شبکه‌های نسل چهارم، شبکه‌های تلفن

همراه نسل پنجم طیف وسیع‌تری از ابزارهـای ارتباطی را هدف قرار داده است. به عنوان مثال، در نسل پـنجم، ایجـاد ارتباط برای دستگاه‌های هوشمند کـه دارای محـدودیت در زمینه انرژی، هزینه و پیچیدگی محاسباتی می‌باشند، یکی از اهداف اصلی می‌باشد [2].

بر اساس یک چشـم‌انـداز بلنـد مـدت، انتظـار مـی‌رود کـه شبکه‌های نسل پنجم و پس از آن، امکان اتصال در مقیاس گسترده، با هزینـه کـم و قابلیت اطمینـان بـالا بـرای تمـام دستگاه‌هایی که قابلیت اتصال به شبکه را دارند، فراهم کنند. تا به امروز، طراحی و بهینه‌سازی شبکه‌هـای ارتباطی و بـه طبع آن دستیابی به درک درست از محـدودیت‌هـای فیزیکـی مانند نویز و تداخل، مبتنی بر مدل‌های آماری بوده است. از سوی دیگر، دستگاه‌های ارتبـاطی کـاربران، کـه غالبـاً تلفن همراه هوشمند می‌باشند، به طور معمول هر روز شارژ می‌شوند و برای برقراری ارتباطات، نیاز دارند به طور مستمر به کانال ارتباطی با ایستگاه پایه مربوطه گـوش دهند. ایـن در حالی است که ایستگاه‌های پایه وظیفه مـدیریت ارتباطـات، ارسال دستورالعمل‌های ارتباطی و زمانبندی منابع رادیویی را برعهده دارند. با در نظر گرفتن پیچیدگی، گسترده‌گی مقیـاس و ناهمگونی شبکه‌های بی‌سیم نسل آینده، بخصوص بـا در نظر گرفتن ترافیک اینترنت اشیاء، امکـان کنتـرل متمرکـز میلیون‌هـا ارتبـاط، در سـمت شبکه بیش از پیش مشکل می‌نماید [3]. در ادبیات این حوزه برای حل این مشکل دو نوع راه‌حل کلی ارائه شده اسـت : دسـته اول، راه‌حـل‌هـای تکاملی است که تـلاش مـی‌کننـد شبکه‌های سـلولی را بـه نحوی توسعه دهنـد کـه ترافیـک مربـوط بـه دستگاه‌هـای اینترنت اشیاء در بستر شبکه‌های سلولی موجود مانند LTE سرویس‌دهی شود [4,5]. دسته دوم شامل راه‌حل‌هایی مـی‌باشد که بر بازبینی اساسی ارتباطات اشیاء با سرورها، کاهش اساسی حجم سیگنالینگ بین دستگاه‌های ارتباطی کـاربر و شبکه، و کاهش کنترل شـبکه بـر منابع مخابراتی متمرکـز شده‌اند [6]. دسته سوم، که دسترسی آزاد بـه منـابع رادیـویی هم نامیده می‌شود، در سال‌های اخیر بـا اسـتقبال زیـادی از سوی گروه‌های تحقیقاتی، استانداردهای مخـابراتی و حتـی صنعت مواجه شده است. به عنوان مثال‌هایی از ایـن دسـته می‌توان به پروتکل‌های ارتباطی مانند سیگفاکس (SigFox) و لورا (LoRa) اشاره کرد.

فناوری‌های لورا و سیگفاکس دو پروتکل غالـب در زمینـه اینترنت اشیاء، بر روی باندهای فرکانسی بدون مجوز مـی‌باشند. همچنین از ایـن دو فناوری مـی‌تـوان بـه عنـوان دو کاندید اصلی پیاده‌سازی نسل پنجم در مناطق روستایی و حومه شهرها نیز یاد کرد. ایـن فنـاوری‌هـا بـا بهره‌گیری از روش استفاده آزاد از منابع رادیـویی (grant-free access)، که در آن نیاز به جفت‌سازی، هماهنگ‌سـازی و رزرو منـابع رادیویی نمی‌باشد، حجم سیگنالینگ مورد نیاز بـین دستگاه (مثلا دستگـاه اینترنـت اشیاء) و سرورها را بسیار کـاهش داده‌اند. این کاهش سیگنالینگ، منجر به افزایش طـول عمـر باتری دستگاه متصل در یک شبکه لورا یا سیگفاکس می‌شود [7]. در حال حاضر، دسترسـی آزاد بـه منـابع رادیـویی یـک موضوع بحث در استانداردهای مخابراتی نیز بـوده و انتظـار می‌رود در نسـخه‌هـای آینـده اسـتاندارد 3GPP LTE نیـز پیاده‌سازی شود [8].

در حالی که میزان مصرف انـرژی در حالـت دسترسـی آزاد بسیار کمتر از حالت با رزرو قبلی می‌باشد، قابلیت اطمینـان ارتباطات در دسترسـی آزاد بـه علـت تـداخل احتمـالی بسته‌های ارسال یک مشکـل اساسی است [9و10و11]. اندازه‌گیری تداخل در فرکانس 868 مگاهرتز (باند فرکانسی ISM) در [9] ارائه شده است. این نتایج نشان می‌دهـد کـه باند فرکانسی ISM، با توجه به رشد روز افزون فناوری‌های ارتباطی که به صورت همزمان در باند فرکانسی آزاد ISM فعالیت می‌کنند، در برخی زیربانـدها و زمانها دارای تـداخل زیادی می‌باشد که این به نوبه خود به معنی احتمـال بـالای تصادم بسته‌ها در این زیربانـدها است.

در سال‌های اخیـر علاقـه فراوانـی بـه اسـتفاده از ابزارهـای یادگیری ماشین برای توصیف رفتار شبکه‌های مقیاس بزرگ به وجود آمده است [12]. در [13]، یک راه‌حل مبتنی بـر یادگیری تقویتی[1]، در سمت شبکه، بـرای شبکه‌های LTE

---
[1] Reinforcement Learning

که به ترافیک اینترنت اشیاء سرویس می‌دهند، ارائه شده است. در [۱۴]، روش خوشه‌بندی خودسازمان یافته و دسترسی به منابع مخابراتی بر اساس خوشه‌بندی، برای شبکه‌های اینترنت اشیاء مورد بررسی قرار گرفته است. در [۱۵]، استفاده از الگوریتم یادگیری $MAB^2$ برای شبکه‌های اینترنت اشیاء پیشنهاد شده است، که در آن دستگاه‌ها یاد می‌گیرند تا از زیرکانال‌هایی که دارای سطح بالایی از تداخل استاتیک هستند، استفاده نکنند. در [۱۶]، برای کنترل زمانبندی و انتقال اطلاعات[3] دستگاه‌های اینترنت اشیاء در سیستم‌های دارای پردازش لبه[4] از یادگیری تقویتی عمیق استفاده شده است. راهحل پیشنهادی در این مقاله قادر است به صورت همزمان مسئله زمانبندی و انتقال اطلاعات را حل کند تا بدین وسیله میزان مصرف انرژی و میانگین تأخیر را به صورت قابل توجهی کاهش داده باشد. در [۱۷]، جنبه‌های مختلف استفاده از یادگیری تقویتی عمیق در برای اینترنت اشیاء خودمختار[5]، شامل مدل‌ها، کاربردها و چالش‌ها، مورد بررسی قرار گرفته است. در [۱۸] تخصیص کانال به صورت توزیع شده و تصادفی بین کاربران شبکه لورا مورد بررسی قرار گرفته است. در [۱۹ و ۲۰]، تخصیص کانال در شبکه‌های اینترنت اشیاء با استفاده از یادگیری ماشینی مورد بررسی قرار گرفته، اما به مصرف انرژی در انتخاب پارامترهای مخابراتی توجهی نشده است.

به منظور تحقق شبکه‌های اینترنت اشیاء خودسازمان‌یافته، که قادر به انطباق با محیط پیرامون خود هستند، ما در این پژوهش سناریوهایی که در آن اینترنت اشیاء بر روی باند با دسترسی آزاد و با حضور شبکه‌های تداخلی دیگر پیاده‌سازی می‌شود، را مورد بررسی قرار داده‌ایم. در این نوع ارتباطات انتخاب پارامترهایی مانند نرخ داده، زیرکانال مورد استفاده، توان ارسالی و تعداد تکرار ارسال، مشخص‌کننده‌ی ظرفیت شبکه و طول عمرباتری اشیاء می‌باشند. هدف اصلی از این پژوهش افزایش سطح اطمینان ارتباطات در دستگاه‌های اینترنت اشیاء ارزان قیمت در محیط‌های تداخلی و کاهش میزان مصرف انرژی در این دستگاه‌ها است. در این راستا، ما یک راه‌حل مبتنی بر یادگیری ماشینی توزیع‌شده، برای کنترل پارامترهای مخابراتی، را پیشنهاد می‌کنیم و نتایج حاصله را با نتایج تحلیلی مربوط به مسئله بهینه‌سازی متمرکز، مقایسه می‌کنیم.

نتایج ارزیابی نشان دهنده کاهش چشم‌گیر در مصرف انرژی و افزایش احتمال موفقیت در ارسال داده است. نتایج اولیه‌ی این پژوهش در [۲۱] ارائه شده است که شامل استفاده از یادگیری ماشین برای تنظیم پارامترهای مخابراتی تکنولوژی لورا در هنگام پیاده سازی است. در این نسخه ما [۲۱] را توسعه داده و نتایج جدیدی در استفاده از یادگیری ماشینی برای سازگاری شبکه با تداخل مقطعی، یادگیریِ کمک‌شده توسط شبکه و تکنیک انتقال یادگیری ارائه می‌کنیم. نوآوری‌های اصلی این مقاله شامل موارد زیر می‌باشد:

- مدل‌سازی تنظیم پارامترهای مخابراتی اینترنت اشیاء به عنوان یک مسئله‌ی بهینه سازی
- بررسی الگوریتم‌های یادگیری ماشینی قابل کاربرد در بهبود عملکرد شبکه‌های اینترنت اشیاء
- ارائه یک راه‌حل مبتنی بر یادگیری ماشینی با هدف افزایش بهره‌وری انرژی اشیاء و قابلیت اطمینان ارتباطات در شبکه‌های اینترنت اشیاء
- توسعه یک مدل تحلیلی برای ارزیابی عملکرد رویکردهای یادگیری توزیع‌شده با بهره‌گیری از ابزارهای آماری هندسه تصادفی
- ارزیابی قابلیت اطمینان، بهره‌وری انرژی و مصالحه‌ی بین آنها در راه‌حل‌های پیشنهادی و سایر راه‌حل‌های موجود در ادبیات حوزه

ادامه‌ی این مقاله بدین ترتیب سازماندهی شده است: در بخش ۲ سیستم مورد بررسی در مقاله تشریح میشود. بخش ۳ به بررسی مبتنی بر یادگیری ماشینی قابل کاربرد در اینترنت اشیاء می‌پردازد. در بخش ۴، یک الگوریتم مبتنی بر یادگیری ماشینی توزیع شده ارائه شده است و عملکرد آن با روش بهینه‌سازی متمرکز مقایسه شده است. نتایج مربوط به

---

[2] Multi-Arm Bandit
[3] Offloading
[4] Edge processing
[5] Autonomous IoT (AIoT)

شبیه‌سازی در بخش ۵ ارائه شده و نتیجه‌گیری نیز در بخش ۶ ارائه شده است.

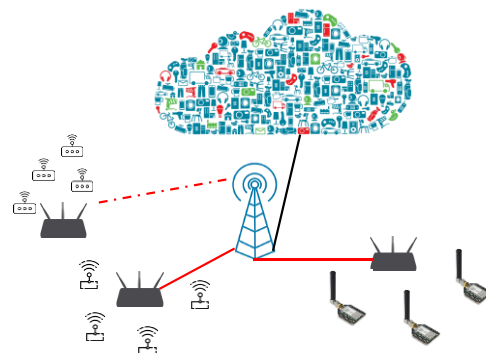

شکل ۱: مدل سیستم، اینترنت اشیاء با دسترسی آزاد

## ۲- مدل سیستم و فرموله‌سازی مساله

سیستم موردنظر در این کار، حالت کلی از ارتباطات اینترنت اشیاء بدون نیاز به دسترسی به کانال، مانند سیگفاکس و لورا، را مدل می‌کند. مجموعه‌ای از دستگاه‌های اینترنت اشیاء، مجموعه $\varphi$، را در یک محدوده جغرافیایی A در نظر بگیرید. انتقال داده از اشیاء به سرور از طریق نقاط دسترسی انجام میگیرد که در این محدوده قرار داده شده‌اند (شکل ۱). هرگاه یکی از اشیاء پاکتی برای ارسال دارد، بدون نیاز به رزرو منبع و زمان‌بندی، اقدام به ارسال پاکت می‌کند. هرنقطه‌ی دسترسی نیز این دستگاه‌های مختلف در شبکه اینترنت اشیاء، دارای الگوهای متفاوت در استفاده از منابع رادیویی می‌باشند، به طور دقیق‌تر، مدت زمان بین دو ارسال داده، پهنای باند سیگنال مورد استفاده، توان ارسالی، نرخ ارسال داده و زمان انتقال بسته‌ها از دستگاهی به دستگاه دیگر متفاوت است. در این مقاله، پهنای باند فرکانسی به اشتراک گذاشته برای ارتباطات را با W و چگالی طیفی توان نویز را با N نمایش می دهیم.

فرض کنید از مجموعه‌ی $\varphi$، زیر مجموعه‌ی $\Phi_s$ موردعلاقه‌ی ما برای جمع‌آوری اطلاعات هستند و ترافیک سایر اشیاء تداخل محسوب می‌شوند. مسئله مورد بررسی در این مقاله، کنترل پارامترهای مخابراتی برای مجموعه‌ای از دستگاه‌های اینترنت اشیاء، $\Phi_s$، بر اساس مشاهده و تعامل با محیط اطراف می‌باشد. فرض کنید در زمان $t$، $i$امین دستگاه از مجموعه $\Phi_s$ نیاز به ارسال داده داشته باشد، در این حالت مسئله موردنظر به صورت زیر فرموله می‌شود:

$$\underset{p_i, r_i, h_i, m_i, \forall i \in \Phi_s}{\text{maximize}} \quad F(\text{REL}_i, \text{EC}_i) \quad (1)$$
$$s.t: p_i \in \mathbb{P}, r_i \in \mathbb{R}, h_i \in \mathbb{H}, m_i \in \mathbb{M},$$

که در آن $F(.)$ بیانگر تابع هدف بوده و تعادلی بین مصرف انرژی (EC) و قابلیت اطمینان (REL) ارتباطات ایجاد می‌کند. با توجه به تفاوت کیفیت سرویس (QoS) در کاربردهای مختلف اینترنت اشیاء، تعریف تابع $F(.)$ در هر کاربرد می‌تواند متفاوت از کاربردهای دیگر باشد. در این پژوهش، مجموع مقادیرِ وزن‌دار و استاندارد شده به بازه‌ی $(0,1)$ قابلیت اطمینان و مصرف انرژی را به عنوان تابع هدف در نظر میگیریم، یعنی

$$F(\text{Rel}_i, \text{EE}_i) = (1 - \beta)\text{Rel}_i + \beta \text{EE}_i, \quad (2)$$

که در آن $0 \leq \beta \leq 1$ پارامتر تنظیم کننده‌ی تعادل بین سطح اطمینان و مصرف انرژی است. همچنین، $p_i$، $r_i$، $h_i$ و $m_i$ به ترتیب اشاره به توان ارسالی، نرخ داده ارسالی، زیرکانال انتخاب شده برای ارسال و تعداد تکرارهای ارسال به ازای هر بسته دارد و نماد $\mathbb{X}$ نیز نشان دهنده مجموعه مقادیر ممکن برای $x_i$ می‌باشد.

یک راهل ابتدایی برای مساله‌ی (۱)، حل آن با کمک ابزارهای بهینه‌سازی و به صورت متمرکز در سرور برای همه‌ی اعضای $\Phi_s$ است. این راه‌حل در کنار اینکه دارای پیچیدگی بالایی می‌باشد نیاز به داشتن اطلاع از پارامترهای مخابراتی سایر اشیاء نیز دارد و در نتیجه در یک رویکرد توزیع شده، قابل کاربرد نیست. بنابراین، به جای استفاده از راه‌حل متمرکز، در این پژوهش از رویکرد یادگیری توزیع شده استفاده می‌شود. در بخش زیر ابتدا یادگیری ماشینی و کاربرد آن در اینترنت اشیاء مورد بررسی قرار میگیرد. سپس در بخش چهارم، به بررسی حل مساله‌ی (۱) با کمک یادگیری ماشینی توزیع‌شده می‌پردازیم.

# ۳- یادگیری ماشینی در اینترنت اشیاء

## ۳-۱- معرفی و دسته‌بندی

یادگیری ماشینی، که در آن اطلاعات موجود توسط یک واحد هوشمند پردازش شده و برای خوشه‌بندی یا پیش‌بینی یا تصمیم گیری به کار میرود، در حالت کلی به سه دسته‌ی یادگیری با داده‌های نشاندار، با داده‌های بی نشان و یادگیری تقویتی[6] تقسیم میشود. در سال‌های اخیر به لطف توسعه قدرت ذخیره و پردازش داده، یادگیری ماشینی در قسمت‌های مختلف شبکه و حتی در سرویس‌های اینترنت اشیاء جای خود را باز کرده است [۲۲]. به عنوان مثال در [۲۳] روش‌های ارائه شده برای امنیت بیشتر ارتباطات اینترنت اشیاء مورد بررسی قرار گرفته‌اند. یک روش پیشنهادی، یادگیری مشخصه‌های فیزیکی هر شیء، مانند تابع توزیع احتمال آفست فرکانسی آن، از طریق بررسی عمیق ارتباطات گذشته و ارزیابی هویت هرپاکت بعدی دریافتی با این مشخصه ها می‌باشد.

بزرگترین مشکل در استفاده از یادگیری با داده‌های نشاندار و با داده‌های بی‌نشان، نیاز به حجم داده‌ای نسبتا بالا برای پردازش و همگرا شدن راه‌حل است، که کاربرد آن را در راه‌حل‌های توزیع شده، در سمت اشیاء، ناممکن می‌سازد و عموما مورد علاقه در سمت شبکه و سرورها می‌باشد. از سوی دیگر، یادگیری تقویتی، که در آن هر شیء از طریق ارتباط با محیط و تجربه، به تدریج پاداش دراز مدت هر عمل را درک میکند، به طور گسترده در سمت ابزارهای هوشمند مورد توجه قرار گرفته است [۲۲]. از مهمترین مدل‌های یادگیری در این حوزه میتوان به یادگیری چند انتخابی(MAB)[7] و یادگیری Q، معروف به Q-learning، اشاره کرد [۲۲]. در هردوی این روش‌ها، کاربر تصمیم گیرنده برای هر عمل یک نمایه در نظر می‌گیرد و بنا به تجربه‌ی حاصل از انتخاب آن، تصمیم به اولویت دادن یا ندادن به آن عمل می‌گیرد. تفاوت این دو الگوریتم، به تفاوت نگاه آنها به محیط برمیگردد که در اولی، محیط با یک حالت (state) مدل می‌شود در حالی که در دومی، محیط پیرامونی با تعدادی حالت مدل می‌شود و ارزش هر عمل در هر حالت جداگانه مورد بررسی قرار میگیرد.

در این پژوهش که اینترنت اشیاء با نیاز مبرم به بهینه‌گی انرژی مورد بررسی ما قرار دارد، تمرکز را بـ ررروی مـدل یادگیری MAB می‌گذاریم تا سربار انرژی و پیچیدگی پردازش داده و تصمیم گیری در اشیاء را به حداقل برسانیم.

## ۳-۲- خود سازماندهی با MAB

در این بخش، ما راه‌حلی مبتنی بر مدل MAB برای خودسازماندهی اشیاء متصل به اینترنت ارائه می‌دهیم. در روش یادگیری توزیع‌شده MAB، هر دستگاه برای بیشینه‌سازی مقدار تابع هدف خود، $F(REL_i, EC_i)$، تلاش می‌کند. این کار با استفاده از بهترین انتخاب ($A_i$) از مجموعه تصمیمات ($\mathbb{A}$) انجام میگیرد: $A_i = \{p_i, c_i, h_i, m_i\} \in \mathbb{A}$. برای انتخاب بهترین تصمیم، پاداش[8] (نتیجه) اقدامات قبلی باید به نحوی ذخیره و پردازش شوند. به عنوان مثال، بعد از انتخاب یک تصمیم در زمان t، یعنی $A_i(t)$، دستگاه میتواند گزارش دریافت از سوی نقطه‌ی دسترسی یا سرور را به عنوان یک پاداش در نظر بگیرد. این گزارش دریافت با $\xi(t) \in \{1,0\}$ نمایش داده می‌شود و در آن ۰ و ۱ به ترتیب بیانگر پیام تصدیق  دریافت و عدم‌تصدیق[9] می‌باشند. هدف عامل بیشینه‌کردن میزان مجموع پاداش در طول زمان و کمینه‌سازی میزان ضررهای[10] متحمل شده در تصمیم‌گیری هایش است. رویکرد مبتنی بر MAB باعث ایجاد تعادل بین فرآیند اکتشاف و استخراج[11] می‌شود. "اکتشاف" به بازه‌های تصمیم‌گیری اشاره دارد که در آن عامل تلاش می‌کند مجموعه گزینه‌های گوناگون را تست کند، حتی در صورتی که پاداش‌های قبلی این اقدامات کمتر از دیگر مجموعه اقدامات باشد. "استخراج" بیانگر زمان‌های تصمیم گیری است که عامل بر اساس مشاهدات قبلی، به صورت حریصانه‌ای برای افزایش پاداش، تلاش می‌کند. با

---

[6] Supervised/unsupervised, reinforcement learning  
[7] Multi-arm bandit  
[8] Rewards  
[9] ACK and NACK  
[10] Regret  
[11] Exploration and exploitation

با توجه به کاربردهای گسترده الگوریتم یادگیری MAB در زمینه‌های مختلفی مانند رباتیک، این الگوریتم به خوبی در ادبیات این پژوهش مورد بررسی قرار گرفته است و راه‌حل‌های کارآمدی به منظور کمینه‌سازی میزان ضرر در تصمیم‌گیری‌ها، پیشنهاد شده است. در ادامه مقاله تلاش شده است تا رویکردهایی برای حل مسئله کنترل پارامترهای مخابراتی در دستگاه‌های اینترنت اشیاء در دو محیط متفاوت ارائه شود: ۱) محیط با تداخل بر روی کانال داده و عدم تداخل بر روی کانال پسخورد، ۲) محیط با تداخل بر روی کانال داده و پسخورد.

## ۳-۳- محیط با تداخل روی کانال داده

در محیط با تداخل روی کانال داده، مساله‌ی یادگیری را MAB تصادفی (stochastic MAB) می‌نامیم. برای MAB تصادفی، که در آن پاداش هر اقدام از یک تابع چگالی احتمال بدست می‌آید، خانواده‌ی الگوریتم Upper Confidence Bound (UCB) میتواند کمینه‌ی ضرر ممکن را بدست آورد [۲۲]. به عبارت دیگر، این الگوریتم به انتخاب تصمیم با بالاترین کران اطمینان در برآورده کردن پاداش بیشینه همگرا می‌شود. در میان الگوریتم‌های خانواده‌ی UCB، ما بر $UCB_1$، ارائه شده در [۲۴]، تمرکز می‌کنیم که در آن ضرر ناشی از تصمیم گیری با نرخ O(log n) رشد می‌کند [۲۵].

## ۳-۴- محیط با تداخل روی کانال داده و پسخورد

در محیط با تداخل روی کانال داده و پسخورد با شرایطی مواجه هستیم که نه تنها یک پاکت ممکن است به مقصد نرسد که ACK مربوط به یک پاکت رسیده به مقصد نیز ممکن است در اثر تداخل به گیرنده نرسد. در این شرایط مساله‌ی یادگیری را به نام MAB غیر تصادفی می‌شناسیم که در آن، پاداش هر تصمیم لزوما از یک تابع چگالی احتمال خاص مشتق نمی‌شود (چون پیام حاوی پاداش نیز ممکن است در اثر تداخل از بین برود). به عنوان نمونه‌هایی از MAB غیر تصادفی می‌توان به محیط با تداخل خصمانه اشاره کرد که در آن یک عنصر متخاصم می‌تواند با تداخل روی کانال پسخورد، کارایی ارتباطات را دچار مشکل کند.

علاوه بر این، در کاربردهای اینترنت اشیاء در باند ISM، تداخل از سوی شبکه‌های دیگر می‌تواند باعث ایجاد تداخل در کانال پسخورد شود.

الگوریتم‌های متعدد برای یادگیری MAB با تداخل غیر تصادفی در ادبیات این حوزه حضور دارند که در میان آنها عملکرد الگوریتم EXP3 دارای عملکرد بهینه می‌باشد [۲۶]. الگوریتم EXP3 در هر زمان تصمیم‌گیری $t$، یک تصمیم از مجموعه $\mathbb{A}$ را بر اساس توزیع‌های احتمال پاداش آنها انتخاب می‌کند، یعنی به بیان آماری: $A_i(t) \sim \{p_1(t), \ldots, p_{|\mathbb{A}|}(t)\}$. میزان ضرر از تصمیم گیری براساس الگوریتم EXP3 با مرتبه زمانی $O(\sqrt{n})$ رشد می‌کند که این میزان، کمینه‌ی رشد در بین سایر الگوریتم‌ها می‌باشد [۲۳].

## ۴- الگوریتم‌های پیشنهادی

به یاد بیاوریم که هدف مسئله بهینه‌سازی (۱)، بیشینه کردن سطح اطمینان و بهره‌وری انرژی دستگاه‌ها است. در این بخش و با هدف ارائه یک راه‌حل توزیع شده برای این مسئله، ما دو الگوریتم یادگیری مبتنی بر MAB ارائه میدهیم که در آنها دو معیار میزان موفقیت و مصرف انرژی در فرآیند یادگیری در محیط با تداخل روی کانال داده و پسخورد لحاظ شده است. این دو الگوریتم در الگوریتم ۱ و ۲ ارائه شده اند. در الگوریتم ۱، در پایان هر انتقال موفقیت‌آمیز داده، فرستنده پیام تصدیق (ACK) دریافت می‌کند و مادامی که این تصدیق دریافت می‌شود، مقدار پاداش تجمعی برای تصمیم مربوطه یک واحد افزایش پیدا می‌کند [۲۲]. در نشانه‌گذاری استفاده شده، $E_k$ بیانگر مقدار انرژی مصرف شده برای ارسال بسته با استفاده از اقدام $k$، $E_{min}$ حداقل انرژی مصرف شده در میان اقداماتی که با موفقیت یک بسته را ارسال کرده‌اند، $T_k(t)$ تعداد دفعاتیکه تا زمان $t$ تصمیم $k$ انتخاب شده است، $A_k(t)$ تصمیم گرفته شده در زمان تصمیم‌گیری $t$، و

$$\bar{z}(t) = z(t)\left[(1-\beta) + \frac{\beta E_k}{E_{min}}\right] \quad (3)$$

نشان‌دهنده‌ی پاداش تغییریافته‌ی تصمیم انتخابی در زمان $t$ می‌باشد که خود تابعی از دریافت یا عدم دریافت ACK

یعنی $z(t) \in \{0,1\}$ است و در ادامه مورد بررسی قرار میگیرد. همچنین، $Z_k(t)$ پاداش تجمعی تصمیم $k$ تا زمان $t$، $\beta$ پارامتر ایجاد تعادل بین سطح اطمینان و بهره‌وری انرژی و در نهایت $I_k(t)$ ارزش انتخاب تصمیم $k$ در زمان $t$ می‌باشد. همچنین، با استفاده از رویکرد مشابهی، در الگوریتم ۲ ما به جای ذخیره و بروزرسانی ارزش یک تصمیم در پارامتر $I_k(t)$، احتمال رسیدن به بیشترین پاداش از طریق تصمیم $k$، یعنی $p_k(t)$ را ذخیره و بروزرسانی میکنیم. در این الگوریتم‌ها، $\alpha$ و $\rho$ پارامترهای طراحی هستند که میان فازهای اکتشاف و استخراج در الگوریتم ۱ و ۲ (به ترتیب) تعادل ایجاد می‌کنند. با مقایسه‌ی الگوریتم ۱ و ۲ با مساله‌ی (۱) مشاهده میشود که $F(\text{Rel}_i, \text{EE}_i)$ در این الگوریتم‌ها با استفاده از تابع پاداش تغییر یافته، یعنی $\bar{z}(t)$ نشان داده شده در (۳)، مدلسازی شده است. در رابطه (۳)، $z(t)(1-\beta)$ نشان‌دهنده‌ی پاداش خارجی (از طرف گیرنده بنا به دریافت یا عدم دریافت پاکت) و $z(t)\frac{\beta E_k}{E_{min}}$ نشان‌دهنده‌ی پاداش داخلی بنا به میزان مصرف انرژی میباشند.

---

**الگوریتم ۱: یادگیری در محیط با تداخل روی کانال داده**

**1** Initialization: $Z_k(1)=0, T_k(1)=1, \forall k \in \mathbb{A}$;
**2** for $t = 1, 2, \cdots$ **do**
  - Update actions' values:
    $I_k(t) = Z_k(t) + \sqrt{\alpha \log(t)/T_k(t)}$;
  - Take the best action: $\arg\max_{k \in \mathbb{A}} b_k(t) \to A(t)$;
  - Update reward counter:
    $Z_{A(t)}(t+1) = Z_{A(t)}(t) + \hat{z}(t)$;
  - Update visit counter: $T_{A(t)}(t+1) = T_{A(t)}(t)+1$ -
  **return** $A(t)$;

---

**الگوریتم ۲: یادگیری در محیط با تداخل روی کانال داده و پسخورد**

**1** Initialization: $W_k(1) = 1, \forall k \in \mathbb{A}$;
**2** for $t = 1, 2, \cdots$ **do**
  - Define distributions:
    $p_k(t) = (1-\rho)\frac{W_k(t)}{\sum_{j=1}^{|\mathbb{A}|} W_j(t)} + \frac{\rho}{|\mathbb{A}|}, \forall k \in \mathbb{A}$;
  - Take the best action: $A(t) \sim \{p_1(t), \cdots, p_{|\mathbb{A}|}(t)\}$;
  - Update the weight function:
    $W_{A(t)}(t+1) = W_{A(t)}(t) \exp(\frac{\rho \hat{\xi}(t)}{|\mathbb{A}| p_{A(t)}(t)})$;
  - **return** $A(t)$;

---

### ۴-۱- انتقال یادگیری

راه‌حل ارائه شده در الگوریتم ۱ و ۲ یک راه‌حل توزیع شده است که بر اساس آن، هر دستگاه پارامترهای مخابراتی خود را با محیط تطبیق می‌دهد. در این قسمت ما به بررسی تکنیک انتقال یادگیری (transfer learning) برای افزایش سرعت همگرایی و در نتیجه بهبود عملکرد سیستم می‌پردازیم. به یاد بیاوریم که الگوریتم ۱ در هنگام آماده سازی، ارزش هر تصمیم را برابر صفر در نظر میگیرد و الگوریتم ۲ نیز احتمال گرفتن بیشترین پاداش از انتخاب هر تصمیم را برابر با سایر تصمیم‌ها می‌گیرد. حال اگر یک دستگاه جدید که شروع به فعالیت میکند از دستگاه‌های اطراف یا نقطه‌دسترسی یا از سابقه‌ی ارتباطات قبلی، ارزش تصمیم‌ها را از یک مقدار اولیه آغاز کند، به این فرایند، انتقال یادگیری میگوییم. انتقال یادگیری، چه از سمت دستگاه‌های قوی‌تر همسایه و چه از سمت شبکه می‌تواند سرعت همگرایی الگوریتم یادگیری را افزایش زیادی دهد که این افزایش با افزایش تعداد تصمیم‌ها قابل ملاحظه خواهد بود.

در بخش بعدی، رویکرد یادگیری پیشنهاد شده برای کنترل اقدامات در دستگاه‌های اینترنت اشیا که به وسیله فناوری لورا به یکدیگر متصل شده‌اند، مورد بررسی قرار گرفته و نتایج حاصله با رویکرد بهینه متمرکز مقایسه شده است.

## ۵- مطالعه‌ی موردی: کاربرد در شبکه‌ی لورا

لورا که به عنوان لایه فیزیکی پروتکل لوراون[12] شناخته می‌شود، تلاش می‌کند که ارتباطات بی‌سیم در فواصل طولانی، با نرخ داده و مصرف انرژی پایین را میسر سازد. ارتباطات در لورا در سه زیرکانال با پهنای باند هر کدام ۱۲۵ کیلوهرتز، و در باند فرکانسی ISM صورت می‌گیرد. مقاومت بالا در برابر نویز و تداخل از ملزومات اساسی برای کار در باند فرکانسی ISM است. بر این اساس، روش مدولاسیون CSS[13] در فناوری لورا مورد استفاده قرار گرفته است. مدولاسیون CCC امکان دریافت و تشخیص سیگنال‌هایی با فاکتورهای پخش (SF) متفاوت، به صورت همزمان را ممکن می‌سازد. فاکتور پخش در لورا بین ۷ تا ۱۲ قابل تنظیم است، که این اعداد اشاره به تعداد chirp های استفاده شده برای کد کردن یک بیت دارد. از این‌رو،

---

[12] LoRa Wide Area Network (LoRaWAN)
[13] Chirp Spread Spectrum

نرخ داده برای کد c با استفاده از فرمول زیر محاسبه می‌شود.

$$R(c) = \frac{c \times BW \times \mu}{2^c}, \forall_c \in \mathbb{C} = \{7, \dots, 12\} \quad (4)$$

که در آن $\mu$ نرخ کد، بین صفر و یک، می‌باشد. بر اساس[27و28]،سیگنال به نویز مورد نیاز، برای تشخیص درست سیگنال‌هایی با فاکتور پخش {7,...,12}، به ترتیب برابر با

$$\gamma_{th}(c) = \{-6, -9, -12, -15, -17.5, -20\}$$

می‌باشد. علاوه براین، با افزایش c، میزان نرخ داده کاهش پیدا کرده و تداخل نسبت به نویز از کاهش نرخ ارسال داده حاصل میشود. ذکر این نکته قابل توجه است که در لوراون برای ارتباطات از توان‌های انتقال 2 و 5 و 8 و 11 و 14 (dBm) را پشتیبانی می‌شود[27و28].

## 5-1- کنترل پارامترهای مخابراتی در لورا

یک شبکه‌ی لورا متشکل از یک نقطه‌ی دسترسی و تعدادی دستگاه متصل را فرض کنید که در آن دستگاه‌ها بر اساس فرآیند نقطه‌ای پواسن[14] (PPP) و با تراکم $\lambda$ در یک محدوده پراکنده شده‌اند. هر فرستنده‌ی لورا به طور میانگین در هر $T_{rep}$ ثانیه یک پاکت به گیرنده ارسال می‌کند. با در نظر گرفتن رابطه (1)، برای تنظیم پارامترهای ارتباطی هر فرستنده، مساله‌ی زیر باید حل شود:

$$\max_{\{c_i, h_i, m_i, p_i\}} F(Rel_i, EE_i) \quad (5)$$
$$\text{s.t:} \quad p_i \in \{2,5,8,11,14\}\text{dBm}$$
$$c_i \in \{7,8,9,10,11,12\}, h_i \in \{1,2,3\}$$

### 5-1-1- رویکرد توزیع شده

با بکارگیری الگوریتم‌های (1)، ارائه شده در بخش4، می‌توان مسئله بهینه‌سازی ارائه شده در رابطه (5) را حل کرد. بر این اساس، مجموعه $\mathbb{A}$ شامل 90 جفت تصمیم خواهد بود که هر جفت شامل یک توان ارسال، یک زیرکانال و یک کد است. نتایج استفاده از رویکرد توزیع شده در شکل 2

---

[14] Poisson point process (PPP)

نشان داده شده است که در ادامه به تحلیل آن خواهیم پرداخت.

### 5-1-2- رویکرد متمرکز

در این قسمت ما به تخصیص بهینه‌ی کدهای ارسال $c \in \{1, \dots, 6\}$ بین فرستنده‌ها می‌پردازیم و تعداد سطح‌های توان ارسالی، تعداد بازارسال و تعداد زیرکانال را برابر 1 می‌گیریم. علاوه بر این، ما تخصیص کدهای ارسال شش گانه را تنها نسبت به تداخل از سوی فرستنده‌های لورا بهینه می‌کنیم و سایر منابع تداخل در باند ISM را درنظر نمی‌گیریم. با در نظر گرفتن این امر که با افزایش کد ارسال از 6 به 12، نرخ ارسال داده کاهش می‌یابد و احتمال تصادم و مقاومت در برابر نویز افزایش پیدا می‌کند، انتظار می‌رود که فرستنده‌های نزدیک به ایستگاه پایه مقادیر فاکتور پخش کوچکتری نسبت به فرستنده‌های دورتر انتخاب کنند و برعکس[11]. بر این اساس، می‌توان ادعا کرد که مسئله تخصیص کدارسال معادل مسئله‌ی پیدا کردن چگالی بهینه فرستنده‌های دارای یک کد ارسال مشابه در هرنقطه از شبکه است. به منظور حل این مسئله در یک فضای دوبعدی، ناحیه سرویس‌دهی را مجموعه‌ای از 6 حلقه تقسیم می‌کنیم که هر حلقه دارای شعاع داخلی $r_{j,1}$ و شعاع خارجی $r_{j,2}$ می‌باشد و اختصاص به یک کد ارسال دارد. با در نظر گرفتن نتایج بدست آمده در [29] و توسعه آنها، می‌توان تابع لاپلاس تداخل مربوط به دستگاه‌های توزیع‌شده در $j$ مین حلقه، که با $\Phi_{j,c} \subset \Phi$ نشان داده شده، را بدین صورت بدست آوریم:

$$L_{\Phi_{j,c}}(s) = \exp\left(-2\pi \int_{r_{j,1}}^{r_{j,2}} \frac{\lambda_{j,c} T_c / T_{rep}}{\frac{1}{sP_t Gr^{-\delta}} + 1} r dr\right),$$
(6)

که در آن $p_t$، $T_{rep}$، $D$، $C$، $T(c) = \frac{D}{R(c)}$ به ترتیب اشاره به توان ارسالی، زمان بین دو ارسال پاکت، طول بسته (بیت)، کد انتخابی برای ارسال و زمان انتقال دارند و $Gr^{-\delta}$ نشانگر افت مسیر است. اکنون می‌توان تابع لاپلاس تداخل دریافتی از همه دستگاه‌هایی که از فاکتور پخش $c$ استفاده می‌کنند را به صورت $\mathcal{L}_{\Phi_c}(s) = \prod_{j \in \mathbb{J}} \mathcal{L}_{\Phi_{j,c}}(s)$ نوشت. اگر نویز سیستم را با $N$ و تداخل دریافتی از سوی

فرستنده‌های گروه $\phi$ را با $I_\Phi$ نشان دهیم، احتمـال موفقیـت در ارسال بسته برای یک فرستنده که از کد $c$ استفاده می‌کنـد و در فاصله $z$ از ایستگاه پایه قرار گرفته است، بـه صـورت زیر بدست می‌آید:

$$p_s(c,z)=\Pr(P_t hGz^{-\delta} \geq \gamma_c N)\Pr(P_t hGz^{-\delta} \geq \gamma_{th_I} I_{\Phi_c}),$$
$$=L_{\Phi_c}(s)\Big|_{s=\frac{\gamma_c}{P_t G z^{-\delta}}} L_N(s)\Big|_{s=\frac{\gamma_{th_I}}{P_t G z^{-\delta}}},$$
$$=\prod_{j\in J}\exp\Big(\int_{r_{j,1}}^{r_{j,2}}\frac{-2\pi\lambda_{j,c}T_c/T_{rep}}{1+(\frac{r}{z})^\delta\frac{1}{\gamma_{th_I}}}rdr\Big)\exp\Big(-\frac{N\gamma_c(c)z^\delta}{P_t G}\Big),$$
$$=\prod_{j\in J}\exp\Big(-\lambda_{j,c}\frac{T_c}{T_{rep}}[Q(r_{j,2})-Q(r_{j,1})]\Big)\exp\Big(-\frac{N\gamma_c z^\delta}{P_t G}\Big)$$
(7)

که $\gamma_c = \gamma_{th_N}(c)$ برابر سطح آسـتانه نـویز قابـل تحمـل برای کد $c$ و $\gamma_{th_I}$ برابر سطح آستانه تـداخل قابـل تحمـل اســت. همچنــین بــرای $\delta = 4$ داریــم:

$$Q(x) = \pi\operatorname{atan}(\gamma_{th_I}^{-.05}\frac{x^2}{z^2})\sqrt{\gamma_{th_I}}z^2. \quad (8)$$

حال می‌توان مسئله (۳) را به صورت زیر بازنویسی کرد:

$$\max_{\lambda_{jc}} \sum_{j\in J}\sum_{c\in C}\frac{\lambda_{jc}}{\lambda}\left[(1-\beta)\int_{r_{j1}}^{r_{j2}}p_s(c\;z)\,dz + \beta\frac{T_c}{T_c}\right]$$

s.t: $\sum_{c\in\mathbb{C}}\lambda_{j,c} = \lambda.$ (9)

با حل این مسئله بهینه‌سازی، چگالی استفاده از هر کد در محیط به صورت تابعی از فاصله تا گیرنده بدست می‌آید. مشاهده می‌شود که حل مسئله بهینه‌سازی بصورت متمرکـز بسیار پیچیده است حتی زمانی که برای ساده‌سازی آن فرض کردیم که دستگاه‌ها در یک فضای دوبعدی و با فرآینـد نقطه‌ای پوآسن توزیع شده‌اند و تنها یک کانال لورا با یـک سطح توان ارسال و بدون نویز خارجی وجود دارد. در ادامه کارائی راه‌حل بهینه متمرکز با رویکرد یادگیری توزیع شده مقایسه می‌شود.

### ۳-۱-۵- مقایسه‌ی رویکرد متمرکز و توزیع‌شده

شـکل (۲) احتمــال موفقیـت در ارسـال داده بــا رویکــرد توزیع‌شده حاصل از الگوریتم ۱ و نتایج مربوط به استراتژی بهینــه متمرکــز را بــرای شــرایط زیــر نشــان می‌دهــد:
$C = \{7,10\}, \gamma_{th_N} = \{-6,-15\}, P_t = 14$ dBm,
$N_d = 1000, T_{rep} = 200, D = 100$ Bytes.

مقادیر مربوط به دیگـر پارامترهـا در جـدول ۱ آورده شـده است. در شکل (۲)، محور x نشان دهنده نمایه‌ی پاکت‌های ارسال شده است. همان‌طور که در بالا نیـز اشـاره شـد، هـر فرستنده به طور مستقل تصمیم می‌گیرد که پاکت‌هایش را با استفاده از فاکتور پخش ۷ یا ۱۰ ارسال کند. بعـد از تعـداد کمـی ارسـال، مشـهود اسـت نتـایج حاصـله بـا استفاده از یادگیری توزیع‌شده بسیار نزدیک به رویکرد متمرکـز شـده است و ضرر ناشی از تصمیم گیری (regret) بـه صـفر نزدیک می‌شود. باید توجه کرد که یادگیری توزیع‌شده بـا وجود آنکه دارای تفاوت کمی با رویکرد بهینه‌ی متمرکز در احتمال موفقیت دارد اما در عمل باعث بهبود قابـل تـوجهی در افزایش طول عمر باتری خواهد شد. دلیل این امر در این است که با این روش، فرستنده بدون نیاز به گـوش دادن بـه سیگنال‌های کنترلی قادر است پارامترهای ارتباطی دستگاه‌ها را به صورت توزیع‌شده تنظیم کند.

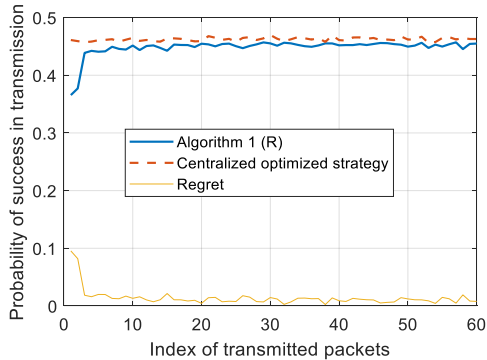

**شکل ۲:** مقایسه الگوریتم ۱ و رویکرد بهینه متمرکز در تخصیص کد به فرستنده‌های لورا

## ۶- ارزیابی عملکرد

در این بخش نتایج شبیه‌سازی عملکرد الگوریتم مبتنی بـر یادگیری ماشینی در فناوری لورا ارائه شده است. در ایـن شبیه‌سازی ۵۰۰ فرستنده در یک محیط دایروی بـا شـعاع ۲ کیلومتر و بطور تصادفی پراکنده شده‌اند. هـدف، توزیـع دو سطح متفاوت توان ارسالی ۸ و ۱۴ (dBm) و انتخاب یـک زیرکانال از ۶ زیرکانال در میان این فرستنده‌هـا می‌باشـد. پارامترهای شبیه‌سازی در جدول ۱ آورده شده اند. در نتایج مربوط به شبیه‌سازی، الگوریتم۱ و الگوریتم۲ اشاره بـه الگوریتم‌های ارائـه شـده در بخـش ۴ دارنـد و الگوریتم ۳

اشاره به الگوریتم متمرکز ارائه شده در [۱۱] دارد کـه در آن بار به طور مساوی بین زیرکانال‌هاتوزیع می‌شود. الگوریتم ۴ معرف الگوریتم مورد استفاده در [۱۸] است که در آن

| پارامتر | مقدار |
|---|---|
| احتمال موفقیت | توزیع شده | تصادفی | 2019 | مرجع [۱۸] |
| احتمال موفقیت | توزیع شده | یادگیری | 2019 | مرجع [۱۹] |
| احتمال موفقیت | توزیع شده | یادگیری | 2020 | مرجع [۲۰] |
| احتمال موفقیت و مصرف انرژی | توزیع شده | یادگیری | 2020 | پیشنهادی |

جدول (۱): پارامترهای ارزیابی عملکرد [۱۶]

| پارامتر | مقدار |
|---|---|
| ناحیه سرویس | دایره‌ای با شعاع ۲ کیلومتر |
| نرخ تجمعی ورود بسته | ۲.۵ بر ثانیه |
| طول بسته | ۲۰ بایت |
| تعداد زیرکانال‌ها | ۱ زیرکانال |
| پهنای باند: $\mathcal{W}$ | ۱۲۵ کیلوهرتز |
| نرخ کد: $\mu$ | ۰.۸ |
| آستانه سیگنال به نویز | $-\{6,9,12,15,17.5,20\}$dB |
| آستانه سیگنال به تداخل | 6 dB |
| توان مصرفی: $P_t, P_c, \eta$ | $\{8,14\}$dBm, 10dBm, 2 |
| پارامترهای یادگیری(پیش‌فرض) $\alpha, \beta, \rho$ | 0.1, 0.4, 0.4 |

زیرکانال بـه صـورت تصـادفی انتخـاب می‌شـود. در پایـان، نمایه‌ی C در جلوی نام یک الگوریتم نشان می‌دهد که فقط کد ارسالی توسط آن الگوریتم انتخاب می‌شود و توان ارسالی برابر ۱۴ dBm هست، در حالی که نمایـه‌ی (C,P) نشـان می‌دهد که توان و کد ارسالی هردو توسط الگوریتم انتخاب می‌شوند. در پایان، الگوریتم ۵ نشـاندهنده روش یـادگیری بکار رفته در [۱۹و۲۰] می‌باشد که برخلاف روش پیشنهادی ما، در تابع پاداش آن صرفه جویی در مصرف انرژی در نظر گرفته نشده است. اما شـکل ۳ احتمـال موفقیـت در ارسال داده را برای ۲ الگوریتم ارائه شده در این کار (الگوریتم ۱ و ۲) و دو الگوریتم پایه برای سنجش (الگوریتم ۳ و ۴) نشان می‌دهد. محور x نشان دهنده‌ی نمایه‌ی پاکت‌های ارسالی (و در واقع نمایانگر سیر زمان) است. در بازه‌ی میانی از زمـان، یک تداخل شدید در ۲ زیرکانال از مجموع ۶ شش زیرکانال اتفاق می‌افتد.

جدول (۲): مقایسه کلی روش‌های بررسی شده در مقاله

| تابع هدف | محل انتخاب | انتخاب پارمتر | سال | نام روش |
|---|---|---|---|---|
| احتمال موفقیت | مرکزی | تقسیم مساوی | 2019 | مرجع [۱۱] |

در شکل ۳، مدت زمان پاسخ به تداخل، طول تداخل و زمان پاسخ به رفع تـداخل مشخص شـده‌اند. مشـاهده می‌شود الگوریتم مبتنی بر یادگیری ماشـینی قـادر است بـا سـرعت خوب خود را با شرایط محیط تطبیق دهد و احتمال موفقیت را در ارسال داده، با وجود و بدون وجود تداخل، بـه مقـدار بیشینه نزدیک کند. از سویی، مشاهده می‌شود الگوریتم ۲، در

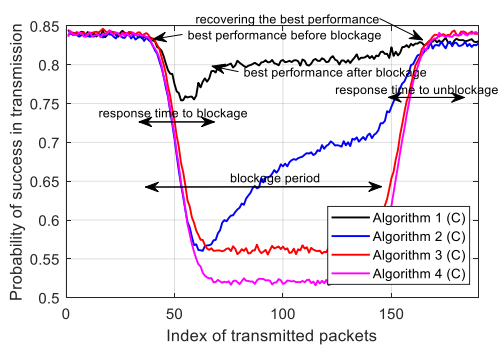

شکل ۳: احتمال موفقیت در ارسال پاکت بر حسب زمان بـرای الگوریتم‌های پیشنهادی و پایه

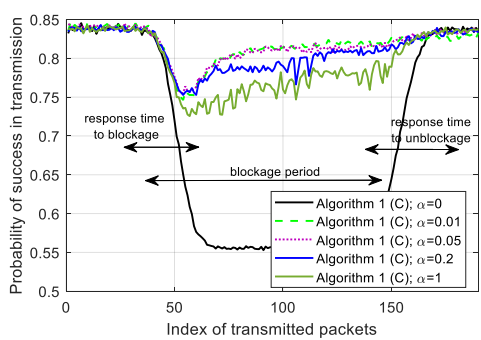

شکل ۴: احتمال موفقیت در ارسال پاکت بر حسب زمان بـرای الگوریتم پیشنهادی با مقادیرمختلف $\alpha$

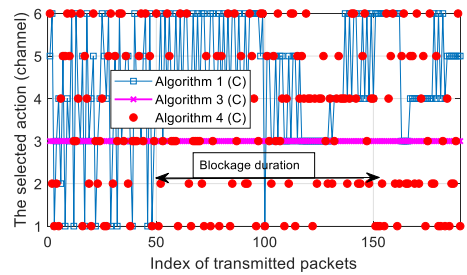

شکل ۵: نمایه‌ی زیرکانال انتخاب شده

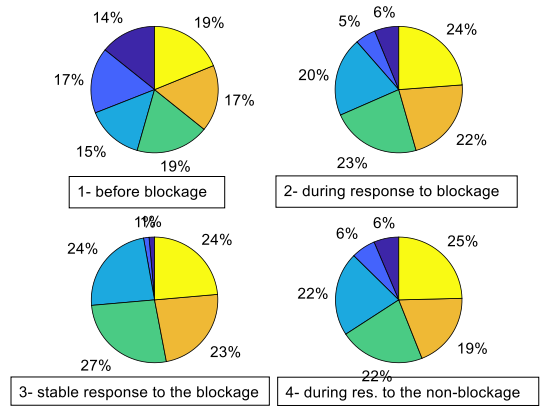

شکل ۶: نحوه‌ی پاسخ الگوریتم ۱ به یک تداخل

این شرایط که روی کانال پسخورد تداخل نداریم، با سرعت پایین همگرا میشود چون به نتایج ACK دریافتی به اندازه الگوریتم ۱ اعتماد ندارد. همین عدم اعتماد است که در

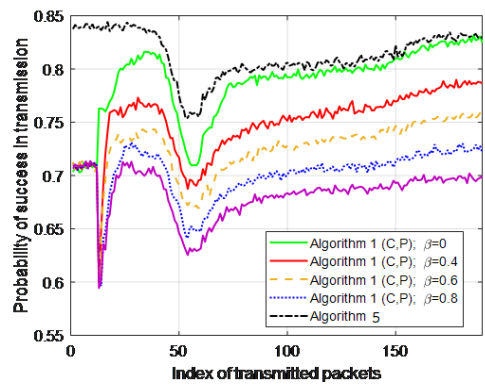

شکل ۷: احتمال موفقیت در ارسال پاکت بر حسب زمان برای الگوریتم پیشنهادی با مقادیرمختلف $\beta$

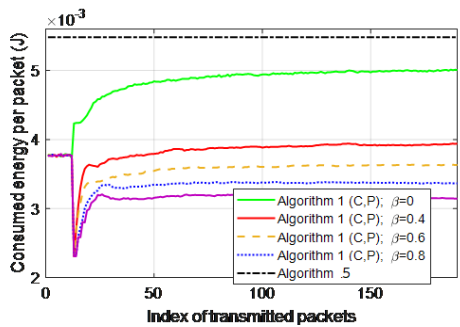

شکل ۸: مصرف انرژی در ارسال یک پاکت بر حسب زمان برای الگوریتم پیشنهادی با مقادیرمختلف $\beta$

شرایط جمینگ که و تداخل روی کانال پسخورد، به این الگوریتم کمک می‌کند عملکرد بهینه‌تری از الگوریتم ۱ داشته باشد [۲۱].

شکل ۴ نشان‌دهنده‌ی عملکرد الگوریتم ۱ برای مقادیر مختلف پارامتر $\alpha$، پارامتر ایجاد تعادل بین تجربه‌ی تصمیم‌های دیگر و استفاده از تجربه‌ی گذشته، است. مشاهده میشود که انتخاب یک مقدار بهینه برای این پارامتر، که در اینجا ۰.۰۵ هست، به سیستم کمک میکند تا هم بتواند در مقابل تداخل‌های احتمال واکنش به موقع نشان دهد و هم در هنگام رفع تداخل به سرعت به حالت بهینه‌ی سابق برگردد.

شکل ۵ نشان‌دهنده‌ی نمایه‌ی تصمیم انتخاب شده توسط الگوریتم ۱ و ۳ و ۴ بر حسب زمان برای یک دستگاه خاص است. مشاهده میشود که دستگاه با استفاده از الگوریتم ۱ پس از تجربه‌ی عدم موفقیت روی زیرکانال ۱ و ۲ دیگر به ندرت به آنها مراجعه میکند.

شکل ۶ نشان‌دهنده‌ی نحوه‌ی توزیع زیرکانال‌ها بین کاربران با استفاده از الگوریتم ۱ در قبل، حین و بلافاصله بعد از رفع تداخل است. مشاهده می‌شود که کاربران قبل از وقوع تداخل روی کانال‌های مختلف بخوبی پخش شده‌اند. پس از وقوع تداخل در دو زیرکانال، استفاده از این دو زیرکانال به تدریج کم میشود. بلافاصله پس از رفع تداخل نیز مشاهده میشود که کابران دوباره به استفاده از این زیرکانال‌ها متمایل میشوند. دقت شود که حتی در میانه‌ی تداخل شدید روی این دو زیر کانال، الگوریتم یادگیری ماشینی آنها را هراز

چندگاهی با ارسال داده مورد بررسی قرار میدهد که این تکرر بررسی با پارامتر $\alpha$ تنظیم میشود.

شکل 7 و 8 نشاندهندهی میزان موفقیت و مصرف انرژی با کمک الگوریتم 1 و الگوریتم 5 در حالتی هست که ما علاوه بر زیرکانال ارسال داده، توان ارسالی را نیز با یادگیری ماشینی تنظیم کنیم. از رابطهی (3) به یاد داریم که $\beta$ تعادل بین بهرهوری انرژی و احتمال موفقیت را در تابع هدف برقرار میکرد. شکل 7 نشان میدهد که با افزایش تعداد تصمیمها (در نظر گرفتن توان ارسالی هم به عنوان یک تصمیم)، زمان لازم برای همگرایی نیز در مقایسه با الگوریتم 5، که تصمیمهای کمتری دربردارد، افزایش مییابد. این تاخیر در همگرایی، در پاسخ الگوریتمها به تداخل ایجاد شده نیز قابل توجه است که نمودار سبز نسبت به سیاه عملکرد پایینتری دارد. شکل 8 مصرف انرژی را برای الگوریتم 1 و هریک از مقادیر $\beta$ درمقابسه با الگوریتم 5 نشان میدهد. مشاهده میشود که استفاده از الگوریتم یادگیری برای انتخاب سطح توان ارسالی همواره به کاهش انرژی مصرفی منجر میشود و این کاهش انرژی با افزایش $\beta$، پارامتر ایجاد تعادل در تابع هدف (3)، افزایش می یابد. مقایسهی تطبیقی شکل 7 و 8 نشان میدهد که کاهش مصرف انرژی ممکن است با کاهش احتمال موفقیت همراه باشد و در نتیجه باید در انتخاب مقدار مناسب برای پارامتر $\beta$ بنا به نوع کاربرد اینترنت اشیاء مورد نظر و کیفیت سرویس درخواستی، دقت لازم بکار برده شود.

## 7- نتیجهگیری

در این پژوهش، راهکارهای ممکن برای بهبود قابلیت اطمینان و طول عمر باتری اینترنت اشیاء مورد بررسی قرار گرفته است. کاهش سیگنالینگ بین فرستندههای اینترنت اشیاء و شبکهی دسترسی در شبکه های ارتباطی مبتنی بر دسترسی آزاد به منابع رادیویی، مانند لورا، منجر به کاهش مصرف انرژی در ارسال دادهها و افزایش احتمال تصادم داده ها در انتقال میشود. در این پژوهش، الگوریتمهای یادگیری ماشینی قابل پیادهسازی در اینترنت اشیاء مورد بررسی قرار گرفته و یک راهحل مبتنی بر یادگیری توزیع‌- شده با پیچیدگی کم برای پیادهسازی در فرستندههای اینترنت اشیاء پیشنهاد کردهایم.

در این راهحل، از پاداش داخلی و خارجی به ترتیب برای کمینهسازی مصرف انرژی و احتمال تصادم در زمان انتقال داده بر روی کانالهای اشتراکی استفاده شده است. سپس عملکرد این روش توزیع شده با روش مبتنی بر راهحل بهینه متمرکز، که با استفاده از هندسه تحلیلی توسعه داده شده است، مقایسه شده و همگرایی الگوریتم توزیع شده مورد تائید قرار گرفته است. نتایج حاصل از شبیهسازی شبکه ارتباطی لورا با کمک الگوریتم توزیع شده نشان از بهبود قابل توجه در احتمال موفقیت در ارسال داده و افزایش طول عمر باتری دستگاهها دارد. این نتایج نشان میدهد که در صورت استفاده از این روش یادگیری در دستگاههای اینترنت اشیاء، این دستگاهها قادر به سازگاری خود با محیط پیرامون و تداخل احتمالی خواهند بود، که نتیجه آن افزایش قابلیت اطمینان در ارتباطات به صورت خودکار میباشد.

## مراجع

# Reliability and Battery Lifetime Improvement for IoT Networks: Challenges and AI-powered solutions

Amin Azari, Mahmoud Abbasi

Towards realizing an intelligent networked society, enabling low-cost low-energy connectivity for things, also known as Internet of Things (IoT), is of crucial importance. While the existing wireless access networks require centralized signaling for managing network resources, this approach is of less interest for future generations of wireless networks due to the energy consumption in such signaling and the expected increase in the number of IoT devices. Then, in this work we investigate leveraging machine learning for distributed control of IoT communications. Towards this end, first we investigate low-complex learning schemes which are applicable to resource-constrained IoT communications. Then, we propose a lightweight learning scheme which enables the IoT devices to adapt their communication parameters to the environment. Further, we investigate analytical expressions presenting performance of a centralized control scheme for adapting communication parameters of IoT devices, and compare the results with the results from the proposed distributed learning approach. The simulation results confirm that the reliability and energy efficiency of IoT communications could be significantly improved by leveraging the proposed learning approach.

Keywords: IoT, 5G, battery lifetime, reliability, machine learning, Multi-arm bandit.